# MAD with aliens? Interstellar deterrence and its implications

Janne M. Korhonen[a]

[a] Aalto University, DF Research Wing, Betonimiehenkuja 5, 02150 Espoo, Finland

janne.m.korhonen@aalto.fi

Tel. +358 41 501 8481

# Abstract

## Abstract

The possibility that extraterrestrial intelligences (ETIs) could be hostile to humanity has been raised as a reason to avoid even trying to contact ETIs. However, there is a distinct shortage of analytical discussion about the risks of an attack, perhaps because of an implicit premise that we cannot analyze the decision making of an alien civilization. This paper argues that we can draw some inferences from the history of the Cold War and nuclear deterrence in order to show that at least some attack scenarios are likely to be exaggerated. In particular, it would seem to be unlikely that the humanity would be attacked simply because it might, some time in the future, present a threat to the ETI. Even if communication proves to be difficult, rational decision-makers should avoid unprovoked attacks, because their success would be very difficult to assure. In general, it seems believable that interstellar conflicts between civilizations would remain rare. The findings advise caution for proposed interstellar missions, however, as starfaring capability itself might be seen as a threat. On the other hand, attempting to contact ETIs seems to be a relatively low-risk strategy: paranoid ETIs must also consider the possibility that the messages are a deception designed to lure out hostile civilizations and preemptively destroy them.

# 1. Introduction

A central fixture in much of the critique expressed against search for and messaging to extraterrestrial intelligences (SETI and METI) has been whether actual contact would be harmful to humanity. In particular, concern has been raised about the possibility of humanity broadcasting its location to hostile extraterrestrial intelligences (ETIs) that might see the Earth as a desirable conquest or humans as a threat to their long-term safety.

Various authors, including both scientists and science fiction writers, have suggested that hostility or fear of hostilities could be a possible or even likely solution to the Fermi Paradox, i.e. lack of contact with ETIs [1–4]. According to these views, the reason why we haven't been able to observe ETIs is either because civilizations are ultimately hostile to each other and contacts between civilizations lead to the destruction of one or another, or because ETIs (in our neighborhood, at the least) believe this to be case and stay quiet for fear of detection. Of these, the former solution is the most unsettling: our lack of caution in broadcasting powerful electromagnetic signals and our efforts to contact ETIs could be inviting an attack.

It should be stated here that I do not argue that such aggressiveness would be the default or even likely attitude for possible ETIs. Without hard evidence either way, it seems at least equally plausible that advanced civilizations (presumably with equally advanced means of destruction at their disposal) have developed ways of life and philosophies that greatly lessen the likelihood of violent conflicts among themselves and with others. Progress towards less violent and more risk-averse societies is evident among humans [5] and simulation experiments, among other things, suggest that cooperative, "pacifistic" civilizations tend to outcompete aggressive ones in the long run [6,7]. Furthermore, the realities of interstellar travel may very well eliminate interstellar aggression as a policy option [8,9]. Obviously, if *all* the ETIs possibly encountered by humans are benevolent, neutral, or unable to harm us, the question whether we would be attacked would be rendered moot.

On the other hand, many rationales for hostility have also been suggested. The non-exhaustive list includes an eagerness to consume our resources, an ideological requirement, a desire to be the sole galactic power, indifference to our existence, or a combination of the above [10]. Science fiction authors in particular have argued that ETIs could see other species as threats to their own well-being, and that paranoid or xenophobic ETIs might simply want to preventively destroy other species before they can pose an intentional or unintentional threat to them [4]. Perhaps the most chilling part of this rationale is that it does not require any particular malice from the part of the ETIs: a simple "rational" if possibly paranoid analysis might suffice to seal our execution warrant. After all, if interstellar travel proves to be feasible, there is always the non-zero probability of humanity harming ETIs.

Prior research has analyzed the general logic and some specific scenarios of interstellar conflict and the risks of contacting possible ETIs (e.g. [7–12]), usually concluding that the practical problems of attacks, invasions and resource grabs would make the possibility of realistic gains dubious at best. However, to my knowledge, no prior work has analyzed the specific case of preventive first strike aimed at eliminating competitors. Although acknowledging the limitations of generalizing from human experience, I believe that the lessons learned from the Cold War – where two fundamentally antagonistic civilizations were able to destroy one another and even had a rationale for launching a disarming first strike – can be fruitfully extended to the specifics of interstellar conflict. In short, this paper tries to roughly estimate the risks of what is perhaps the most disquieting scenario: that an ETI would, upon detecting advanced civilization on Earth, launches an unprovoked preventive attack aimed at destroying or severely damaging the humanity.

While this analysis does not cover irrational attacks (e.g. ones motivated by ideology, such as xenophobia), carelessness or accidents, the findings do suggest that the possibility of retaliation would seem to make preventive attacks a flawed strategy, and that interstellar civilizations would be disinclined to knowingly initiate hostilities against each other in general – even under the worst case assumption that communicating peaceful intentions and working towards win-win solutions across interstellar distances proves to be impossible. In fact, even irrationally aggressive civilizations can probably be deterred. The paper should help to shape the discussion of risks of ETI contact and the findings should help ameliorate one of the objections expressed against METI efforts, namely, that METI would broadcast our location to a possibly hostile ETI (for an overview of the discussion, see e.g. [12,13]). However, the study also advises caution in the design of interstellar exploratory missions.

This paper is organized as follows: I will first briefly revisit the general arguments for and against of what I call "paranoid" attitude towards other civilizations and note the similarities to relevant Cold War arguments. I then discuss the general problems facing a would-be attacker, and the requirements of deterrence. Then, I create a simplified model of interstellar attack and retaliation in order to illustrate the difficulties facing the attacker. Finally, conclusions and a discussion are provided.

## 2. Why ETIs would want to kill us?

In the past, several commentators have noted that we cannot rely on possible ETIs being benevolent towards us. Looking at examples of human and animal behavior towards other species and technologically less advanced cultures (e.g. [14]) and considering that any ETIs would be very likely to be much more advanced than humanity, there is a chance that humanity could be attacked if detected. The proposed motives for attack range from indifference and completely alien (i.e. incomprehensible) reasons to us being seen as a threat, an useful source of food or other resources, or simply as sources of good entertainment [14–16].

Although these concerns cannot be disproved, at least several of these rationales seem *a priori* rather far-fetched reasons for any advanced civilization to pose an existential threat to us. For example, a civilization capable of large-scale interstellar voyages must of necessity possess knowledge and energy reserves sufficient to satisfy most of its material needs via permanently recycling closed-loop economy within its spaceships [17]. If such technologies can be developed for use onboard spaceships, utilizing them within the species' home system should be much more energy efficient and safer than raiding inhabited systems for resources [18].

The exception to this rule, an exponentially expansive civilization could, in theory, populate the entire galaxy in very short timescale while "strip-mining" star systems in the process. However, since we haven't seen any evidence of civilizations that strip-mine everything on their path, the existence of such civilizations in our galaxy at the least seems doubtful [10,18]. It would therefore appear believable that, at least for the foreseeable future (thousands of years at the least), "true" interstellar civilizations would have ample space and would not have to risk antagonizing upcoming species.

Even if they colonize inhabited star systems, the inner planets and their resources may not even be very attractive when compared to resources available in e.g. asteroid belts and gas giants.

One would indeed be hard pressed to understand why any star faring ETI would even bother threatening less advanced civilizations for what can be obtained through less risky and more profitable methods, e.g. cooperation and trade. The economic principle of comparative advantage should hold even between civilizations with immense differences in development, and therefore relations between two species would not be a simple zero-sum game of winners and losers.

Therefore, arguments that any civilization that has truly mastered interstellar travel would have much to gain by destroying non-spacefaring species seem implausible. Destroying a species that cannot harm the invader would not improve the invader's security at all, and the gain of a single planet would seem to be a trivial advantage to a civilization that already has the capability to live in space. Against this small or even negative gain, the extermination attempt risks leaving survivors or witnesses that in all likelihood *would* become security threats in the future.

## 2.1. Small is dangerous: the threat of single-system civilizations and interstellar exploration

However, the situation is somewhat different if the civilization is less advanced than a "true" star faring civilization. Perhaps the most threatening civilization will be the one that is unable or unwilling to colonize other star systems or even its own home system, but still has the capability to explore and industrialize inter-system space. Although such a "single-system" civilization would not threaten other civilizations for living space or resources, it (or, more to the point, its leaders) might feel threatened by the very existence of other civilizations. The reason is fundamentally simple: a civilization living on one or at most a few worlds is inherently vulnerable. The vulnerability is compounded by lack of information about the other civilizations and their intents. If single-system civilization detects another civilization, immense distances and the associated light speed lag will most likely separate them. Moreover, alien cognitive processes are by definition likely to be very difficult to understand. This means that the single-system civilization cannot be at all certain whether or not the other might be planning an attack for some reason.

Even if the other has no hostile intentions, star faring capability in itself can be seen as a threat. Just ordinary interaction between two species could inadvertently destroy or seriously damage one of them, perhaps through transmission of diseases, invasive species, or undesirable information [10]. Furthermore, any spacecraft capable of interstellar voyages in reasonable time is by itself a weapon of mass destruction. Even if "breakthrough physics" concepts such as wormholes and warp drives [19] (all of which could be misused) prove to be impossible, relatively simple interstellar probes – possibly within humanity's capabilities in the relatively near future [20–24] – could be devastating weapons. To illustrate this, Table 1 shows the kinetic energy for each 1000 kg of spacecraft mass at different velocities. For comparison, the largest nuclear weapon exploded on Earth yielded ≈ 0.05-0.06 gigatons and the entire global nuclear stockpile has been estimated at 6.5 Gt.

*Table 1. Kinetic energy of 1000 kg impactor at various velocities, in gigatons of TNT equivalent (1 Gt = 4.184 x $10^{18}$ J)*

| V/c | KE/1000 kg |
|---|---|
| 0.1 | 0.107 Gt |
| 0.25 | 0.709 Gt |
| 0.5 | 3.33 Gt |
| 0.75 | 11 Gt |
| 0.99 | 130.8 Gt |

It is easy to see that even primitive interstellar probes, traveling at an appreciable fraction of light speed, could be extremely dangerous to planet-bound civilizations. Besides intentionally hostile acts, simple accidents or acts of lunacy or recklessness could have extremely grave consequences for other species as well. At current and near-future levels of technology, detecting and intercepting such projectiles in time would be extremely difficult. Compared to objects of scientific interest such as asteroids, hostile probes could be relatively small and have very little apparent motion relative to their targets, effectively hiding them against the background. Although discussion of probe detection and planetary defense are beyond the scope of this paper, it seems to be that reliable detection and defense would require advanced technologies and significant off-planet infrastructure[1].

These considerations may be interpreted to suggest that all that is required for a civilization to pose a mortal threat to another is the *capability* to do harm. Presumably, because something bad *can* happen otherwise, any species (and in particular, vulnerable single-system or single-planet species) should be predisposed to preventive elimination of potentially dangerous adversaries before they have the capability to do the same. Contrary to argument expressed by e.g. Musso [25], destroying another civilizations would *not* necessarily require "very evil species" inspired by "almost satanic will" (p. 51).

---

[1] A rough calculation made by the author suggests that the passive infrared signature from a primitive Daedalus-type [24] heading towards the Earth at 0.12 *c* might in theory be detectable at ≈ 10-20 AU with Hubble Space Telescope-level sensors . Detection even at such ranges would give only some 12 to 24 hours of warning before impact (with total kinetic energy of ≈ 145 Gt). If the probe is targeted on the Earth on purpose, as opposed to e.g. software error or malfunction, it is also reasonable to expect that its design would incorporate some low observability features. These could probably reduce the probe's signature considerably.

It should be noted here that the definition of "single-system" could cover not just the species' original home world, but also its colonies. If interstellar travel and communication remains difficult, individual colonies are likely to be very independent, and therefore are likely to make decisions about e.g. the risks and benefits of waging war with their neighbors independently from the broader considerations of the home world. Similarly, the arguments presented here will apply not just to a contact between two species, but also to relationships between two inhabited star systems, e.g. colonies and home worlds.

## 2.2. Lessons from the Cold War

Readers acquainted with the history of early Cold War and debates about nuclear strategy will find the above argumentation eerily familiar. Even serious early-war studies argued that since nuclear war would be almost certain to happen sooner or later, the U.S. should take the initiative and wage preventive war, on its own terms, against the Soviet Union before it became an existential threat [26,27]. For example, in 1954 a Joint Chiefs of Staff advance study group briefed the president Eisenhower on a plan proposing that the U.S. should "deliberately precipitat[e] war with the USSR in the near future… before the USSR could achieve a large enough thermonuclear capability to be a real menace to [the] Continental U.S." [27] (p. 101) Another contemporary study concluded that anyone calling for restraint and relying on retaliation in the event of nuclear attack was a "pseudo-moralist who insists that [the U.S.] must accept this catastrophe." [27] (p. 100)

Mathematically, the logic was sound. If a war had non-zero probability, given enough time, it would certainly occur. Given the trend towards increasing destructiveness and numbers of nuclear weapons, war in the far future would be more destructive than a war in the near future. Therefore, a logical conclusion would be to choose the least bad from two "tragic but distinguishable post-war states," to use Herman Kahn's memorable description.

## 2.3. Retaliation and deterrence

Needless to say, the "pseudo-moralists" still won the debate. Besides moral issues that troubled even generals (see e.g. [26,27]) and obvious political difficulties, preventive war would not have been reliable by the forces available to the U.S. in the late 1940s. After the Soviet Union got the bomb in 1949, any attempt would almost certainly have resulted to a retaliatory response. Despite any defenses, some warheads would inevitably have gotten through, and losing "even" a single city would have been an unmitigated disaster by any practical measure. As the Cold War progressed, the increasing numbers of nuclear weapons and nearly invulnerable second-strike systems (in particular, submarines), together with the realization that global effects of nuclear war (e.g. fallout and nuclear winter) could doom the humanity, practically ended speculations about intentional, preventive war [28].

In other words, fear of retaliation – i.e. deterrence – has been a major (albeit far from the only) reason why no state has initiated a preventive nuclear attack against another [29]. Although defining what constitutes an effective deterrent is difficult if not impossible, a tacit understanding seems to exist that deterrent is reliable if it can inflict "unacceptable" damage to the attacker. In one recent example [30], the capability to destroy any ten cities in retaliation after a surprise attack was seen as reliable and adequate nuclear deterrent between the U.S. and Russia. Our own experience would therefore suggest that advanced civilizations tend to be risk-averse and do not want to gamble with even small portions of their heartland, unless left with absolutely no other option.

Of course, extrapolating the behavior of ETIs from human examples is risky. The concept of unacceptable damage may simply not apply. Presumably, the inhibitions against killing (if they exist at all) will not be as strong when the target is another species, as can be demonstrated by how humans treat even relatively intelligent, harmless species such as dolphins or chimpanzees. If the biospheres are separated by light years, the concerns about global effects are moot, and it is possible that the technological disparity between the attacker and the defender is so great that the defender cannot effectively retaliate. After all, the age difference may be vast: the median age of terrestrial planets in our galaxy, for example, is likely to be 1.8 billion years more than the Earth's [31].

However, the features of interstellar conflict will also remove some of the major objections raised against Mutual Assured Destruction (MAD) as a deterrent strategy between human civilizations on Earth. For example, in response to an attack aimed at destroying the entire species, the concept of disproportionate response does not seem to be relevant. Likewise, lack of psychological inhibitions and isolation from ecosystem damages will also remove some of the objections [32] and therefore strengthen deterrence by making retaliation more likely. The major drawback of MAD doctrine, the need to keep large nuclear forces on alert and the resultant risk of accidents or sabotage, is almost completely averted: as the flight times to targets are in any case measured in decades or centuries rather than minutes, the time spent for preparing the retaliatory response is not so critical. The remaining question mark would therefore seem to be the technological disparity: can the attacker count on being able to prevent the retaliation?

## 3. General problems of interstellar conflict

In the following, I shall consider certain problems that will be applicable to any interstellar conflict between two civilizations, heretofore termed as Attacking Civilization (AC) and Victim Civilization (VC). The term "civilization" may refer to a single civilization or a group of civilizations either acting jointly or being affected at the same time. I will also consider the implications of *other* civilization(s) not directly involved in the initial conflict, the *N*th Civilization (NC). Before considering the specific problems, the following six major and one minor assumptions about the nature of interstellar conflict are outlined:

1. *All civilizations will have a concept of risks and benefits,* i.e. they are somewhat rational actors and do not simply act randomly.

2. *A civilization that does not need to fear retaliation has little need to destroy other civilizations*. The star faring civilizations discussed above will have sufficient knowledge and resources on their disposal to have no real need to exterminate planet-bound civilizations. Although accidents, carelessness and attacks stemming from e.g. xenophobia or completely alien motives cannot be ruled out, it would seem that deliberate attack aimed at the destruction of an irrelevant species would expend resources to little purpose. After all, as common sense and the Table 2 show, if the VC *cannot* be a threat to the AC, ever, destroying it does not change the security situation of the AC. The AC not attacking would therefore seem to be the optimal strategy. Another reason to doubt the possibility of attack by most ancient civilizations (e.g. $\approx 10^9$ years old) is the fact that such civilizations would have had ample time to explore throughout the galaxy. If such civilizations were to be hostile, they would have also had time to plant automated sentinels on most if not all star systems. Delaying the attack until the VC has developed a technological civilization would seem to be a deeply flawed strategy.

If only a civilization that has a reason to fear the VC's current or future capabilities should have a reason to try to destroy it, then the major question of deterrence in interstellar relationships – whether the technological disparity will always be too great for the VC to effectively threaten with retaliation – has a simple, negative answer.

*Table 2. Simplified outcomes from non-threatening and threatening Victim Civilizations. "Cooperation" refers to non-zero sum cooperation, such as trade.*

|  | VC cannot be a threat | VC can be a threat |
|---|---|---|
| **AC will attack** | No security gain for AC | VC may retaliate |
| **AC will not attack** | No security gain for AC VC may cooperate | VC may attack or cooperate |

3. *There are practical limits to technological development.* Although debatable, for the purposes of this paper it seems reasonable to believe that civilizations will at some point reach a stage where they will not be able to greatly reduce the vulnerability of their habitats and other installations through technological improvements (see also [33]).

Because of the third assumption, I can also assume that *no defense can be guaranteed to be 100% successful 100% of the time.* In other words, there is always a possibility that an attack will slip through even the most elaborate and advanced defenses.

5. *No attack can be guaranteed to be 100% successful.* There is always a chance that any attack will fail to achieve its complete objectives. Particularly if the objective is the total destruction of the Victim Civilization, being completely certain about complete success remains difficult despite technological advancements. For example, self-contained space habitats in the outer reaches of the star system could conceivably escape initial attacks, or the VC might have even sent out colonization ships after the initial detection but before the attack.

The outcome of the last three assumptions is clear. To avoid retaliation, the AC needs to be nearly 100% certain to destroy the VC's ability to retaliate. However, effectively deterring the AC requires only that the VC have the capability to threaten unacceptable damage to the AC. The analogue to modern-day nuclear forces is direct: creating absolutely effective first strike weapons and gathering timely intelligence required for their use will always be massively more complicated than creating effective deterrent weapons.

6. *Verification of peaceful intentions may be difficult.* The interstellar distances make any communications and assurances doubtful; there is little that can be done, short of a physical visit, to verify the truth of any statement any civilization may make. In the worst case, crafting "win-win" strategies and easing tensions may be impossible.

Finally, this study assumes that *the light speed limit holds.* Although this assumption can be relaxed considerably without altering the conclusions unduly, the conclusions are even stronger if matter or information cannot be transmitted faster than light.

## 3.1. The problem of interstellar intelligence gathering

The principal problem facing any would-be Attacking Civilization is knowing what to attack. Solid intelligence has been seen as the essential ingredient of any attack planning among humans, and it is difficult to see that that would be very different with any conceivable aliens. However, gathering that intelligence might be very difficult and any results fundamentally uncertain.

Consider just the simplest problem, finding all the habitats that need to be targeted for destruction. Even at less than interstellar distances, accurately identifying such habitats may be difficult. Unless these can be destroyed, the AC must fear retaliation, as there is always a chance that the VC learns enough about the AC to deduce the origin of the attack and retaliate against AC's home worlds – even if the AC manages to inflict mortal wounds on the VC.

What's more, if the light speed limit holds, all intelligence gathered before an attack is launched will be very much out of date by the time the attacking force arrives to the target system. This is not so much a problem if the attack is simply intended to inflict as much damage as possible, as would be the case with a retaliatory strike. However, it is a serious problem if the objective is the complete destruction of the Victim Civilization, so that it is unable to retaliate any time in the future. In the worst case, if the AC seriously misjudges the VC's speed of development, the attacking force may be outclassed by centuries of technological development, its intentions correctly surmised (perhaps even if it can decide not to attack) and a retaliatory response mounted without any damage to the VC. Besides technological development, the time lag gives the VC more opportunities to establish contacts with other civilizations, maybe simply by sending out colonization ships to other star systems. As seen later, these *N*th Civilizations pose severe problems to the attacker.

The average intelligence lag may be estimated from the estimated average density of advanced civilizations or their colonies within the galaxy. Assuming, for simplicity, the galaxy to be a disk with $r = 50\,000$ *ly* and $h = 1000$ *ly*, even highly optimistic estimates – one million evenly distributed civilizations – suggest that the average distance between civilizations is $\approx 200$ *ly*. Then the minimum intelligence lag seems to be on the order of $\geq 400$ years (200 years for signal, another 200 years for near-*c* travel). With only marginally less optimistic assumptions (100 000 civilizations or colonies), the average spacing would be $\approx 430$ *ly* and the minimum intelligence lag would be almost 900 years. Other authors have reached conclusions suggesting an average spacing between "few hundred" and 1700 light years [34]. If the detection were supposed to happen because of artificial electromagnetic emissions or other signs of technological civilization (e.g. changes in atmospheric composition), this would suggest that human-type civilizations have time to develop at least primitive retaliatory capabilities by the time any adversary can mount an effective attack (see also [9]). Earlier detection is, of course, possible (e.g. from searches for life-bearing planets), but then the question why humanity has not already been eradicated becomes difficult to answer.

### 3.2. The problem of pace of development

Prior literature has repeatedly argued that any extraterrestrial intelligence would be significantly more powerful than human civilization [35]. This is because humans and human technology have been relatively recent phenomena in the history of Earth and the universe, and because technological development seems to happen relatively quickly compared to evolutionary timescales [10]. Given the age of universe and of our galaxy, it would also seem very highly unlikely that two intelligent species would develop technological civilizations exactly at the same time. Consequently, it is argued, if humans encounter any ETIs at all, they are likely to be very highly advanced.

However, this argument cuts both ways. Any given civilization encountering signs of any other civilization for the first time will be fundamentally uncertain as to what is the level of their development. The only thing they will know for sure is the same thing we know now: it is highly unlikely that they are exactly matched.

Although it is usually assumed that the level of technological achievement can be deduced from various signatures of the civilization (electromagnetic radiation, signs of megascale engineering projects, etc.; see e.g. [34]) it is by no means certain that highly advanced technologies have to leave highly visible footprints [36]. We also know from human history that many technologies remain in widespread use long after they have been made obsolete in their primary tasks, if only for entertainment or educational purposes. Additionally, it is far from certain that technological development proceeds at a same pace across different cultures [37]. In short, determining whether observed signatures really represent the genuine capabilities may be extremely challenging, particularly if the observations have to be made across interstellar distances.

Suppose, for example, that an invader from the 16$^{th}$ century approaches a modern coastline resort. Upon seeing the sails of various pleasure craft, he might deduce that our level of technological advancement is not markedly higher than his, and that we pose no threat to his war galleon. Similarly, any ETI who is perhaps listening may have problems knowing for certain whether signals emanating from the Earth are the genuine legacy of technological development, or a hyper-advanced civilization's analogue to pleasure boating or modern-day renaissance fair.

What's more, if an advanced civilization has reasons to believe that other, potentially hostile civilizations exist, it may find "bear-baiting" a very attractive strategy. Like a hunter using live bait to lure out a bear, advanced civilizations might clandestinely follow upcoming civilizations or even create decoys that mimic the signatures of less advanced civilizations in the hopes of drawing a response. The rationale for using such baits is simple: even advanced civilizations may have reasons to be wary of hostile neighbors. Therefore, drawing out hostile civilizations and pre-emptively attacking them could be a prudent strategy.

### 3.3. The problem of survivors

The primary concern for any potential aggressor would be whether or not the hostile acts will be met with a reprisal. Unfortunately, the distances, timescales and cultural differences involved would suggest that making meaningful agreements to end an interstellar conflict and monitoring their compliance in the long term are going to be nearly impossible (see also Assumption 6 above). Therefore, the only sure way to ensure victory would seem to be the complete extermination of the other species.

By default, any overt attack against a civilization proves to that civilization that somewhere out there is a threat capable of severe aggression over interstellar distances. The more severe the initial attack, the more likely it is that the Victim Civilization will deem that its long-term survival will require retaliation.

If the VC survives the attack in any form, it is very difficult for the AC to ensure that it cannot or will not retaliate at some time in the future. Left to their own devices, even a handful of survivors could repopulate entire planets relatively quickly: for example, an average growth rate of 1% - mediocre by historical standards – could repopulate the Earth to seven billion people from only a five thousand survivors in little more than 700 years. If the motive for initial attack had been to ensure long-term survival of the Attacking Civilization by wiping out the competition in the stellar neighborhood, a gain of mere 700 years would constitute a massive failure. Even "knocking a civilization back to Stone Age" might theoretically mean only some tens of thousands of years before the said civilization could pose dangers to the attacker. On galactic timescales, even such a respite is temporary at best.

It is, of course, uncertain whether such survivors would ever be capable of launching a successful retaliatory strike. It is also uncertain whether they would want to do so, if not for any other reason then because any attempts would draw renewed attention to them. However, as long as there are any survivors, the attacker cannot entirely discount the possibility. The question, then, is whether the attack leaves any survivors capable of holding a grudge. Fortunately from our viewpoint, ensuring the complete destruction of an advanced species may be challenging. It is likely that just 100-200 years will suffice to give humanity, for example, the first permanent foothold in space. Unless *all* space habitats and off-world colonies can be destroyed in the attack, it is quite plausible that the survivors will give high priority to hitting back at their attacker. Even without space habitats, complete elimination of the dominant species of a planet would seem to be a relatively uncertain undertaking.

An obvious counter would therefore be that the AC would seek to "occupy" the target system(s) for a long time, perhaps with automated sentinels, in order to mop up the survivors as they are spotted and/or send advance warning of a possible retaliation.

However, this requirement adds significantly to the cost, complexity and uncertainties of the attack.

### 3.4. The problem of witnesses

Furthermore, even if the Attacking Civilization manages to completely eliminate the Victim Civilization, there is always a possibility that *other* civilizations – the *N*th Civilizations or NCs - take notice. Logically, any civilization initiating an unprovoked attack against another would be very dangerous to other civilizations as well; therefore, the AC needs to be sure that other, perhaps so far undetected civilizations do not reply with preemptive attacks. It should be noted that these other civilizations include not just other aliens, but also possible space-faring relatives of the VC's (or even AC's) civilization – e.g. colonies or original home worlds. In fact, given the probable rarity of advanced species in the galaxy, if NCs exist, they are more likely than not going to be off-shoots of the same species as the VC or the AC. Such offshoots would seem to be likely to be in contact with each other, and therefore able to warn others of an attack, even if the attack succeeds. Therefore, the AC needs to be fairly certain that VC is not in contact with other civilizations that have the capability to pre-empt. Admittedly, the risk of pre-emption from NC may not be as large as the risk of retaliation from the VC: as the lightspeed lag applies to the warning messages from VC to NC as well, the AC will at the minimum have more time to prepare for possible retaliation. On the other hand, the capabilities of the NC are more likely to remain unknown to the AC, and the AC has to take this into account as well.

In the end, members of same species might be the greatest danger to a civilization that displays overtly aggressive tendencies. After all, an aggressive relative also poses a threat to *them*, both directly and indirectly. The direct threat stems from the fact that members of the same species are also the closest competitors in terms of habitat requirements etc., and a civilization willing to utterly exterminate aliens may not balk at using force against its own species. The indirect threat may materialize if the aggressive member of the species provokes an interstellar conflict, as the retaliation is unlikely to be very selective.

## 3.5. The problem of learning from potentially lethal experiments

It could be argued that all of the above are inferences based on a sample of one or zero, and that a hyper-advanced civilization might be much better at estimating the strengths and weaknesses of any potential Victim Civilization. But one must consider what this implies: accurate assessment of the strengths and weaknesses of an alien civilization requires experience with alien civilizations. Although one of the traditional assumptions of SETI has been that any ETIs humanity may encounter are likely to have encountered other ETIs before (for an example, see e.g. [12]), a civilization with experience in fighting other civilizations has to be one that has encountered other civilizations in the past, fought conflicts against them, and survived. While it is far from impossible to imagine contacts leading into conflicts, the extreme uncertainties and expenses of interstellar conflict makes it difficult to believe that conflicts between civilizations would be common. It is even more difficult to believe that such conflicts happen so frequently that civilizations would have survived long enough and accumulated enough experience to determine the capabilities of newly contacted civilizations with near-perfect certainty. In fact, the more advanced civilizations there are out there, the more likely it is that any would-be AC would have a problem with $N$th Civilizations or even coalitions of civilizations. Although the immense distances mean that alliances are unlikely to be able to help their individual members in time to fight off a surprise attack, it would still be in the interests of "peaceful" civilizations to gang up against any civilization displaying aggressive tendencies, and as e.g. simulation studies of international relations suggest, aggressive civilizations may be at a long-run disadvantage against peaceful but vigilant cooperatives — even when the retaliation is not immediate [6].

Thus, even very advanced civilizations are in all likelihood quite unaccustomed to interstellar conflict and unlikely to have a history of preventively eliminating other civilizations. Assuming that they would decide to attack another civilizations simply because they *could* be threats in the future would seem to be highly unrealistic.

## 4. Modeling the decision-making of interstellar attack

The considerations stated above must necessarily affect the probability that any civilization decides to attack another civilization. To provide a moderately analytical starting point for the discussion, we can simply calculate the rough probability that the Attacking Civilization will escape retaliation from the Victim Civilization and pre-emption from possible $N$th Civilization(s).

Let the probability $P_{unpunished}$ be the joint probability that the intelligence is adequate and timely enough so that 1) the VC's (and its allies') essential centers of gravity are identified correctly ($P_{identified}$), 2) the attack hits the targets with sufficient force ($P_{hit}$), 3) the Victim Civilization's (and/or its allies') ability to retaliate has been permanently destroyed ($P_{destroyed}$), and 4) there are no $N$th Civilizations capable and willing to preempt the Attacking Civilization ($P_{\neg NC}$). The probability that the AC will be attacked in return will therefore be the complement,

$$P_{punished} = 1 - (P_{identified} \times P_{hit} \times P_{destroyed} \times P_{\neg NC}) \qquad (1)$$

Even if the AC is 95% certain of each individual variable, the probability of counter-attack is uncomfortably high 0.185; if the certainties are "only" 90%, the $P_{punished}$ will be 0.34. I will leave it to the reader to judge whether, given the arguments above, confidence levels of 90% or 95% would be attainable in reality.

Of course, what exact probability of retaliation is required to deter an attack remains a matter of debate. Although the limitations of extrapolating from human experience must be acknowledged, we may be able to draw some insights from the debate about nuclear deterrence. As mentioned above, a recent work assumed that the capability to hit ten cities, with one nuclear weapon each, in retaliation would be a credible deterrent for the U.S. and Russia both [30]. The assumption did not require the largest 10 cities to be hit, but if we assume that the retaliation would hit and completely destroy the largest ten cities in the U.S. (two highly unrealistic assumptions, but ones that increase the margin of error), the retaliation wouldn't need to threaten more than 24.5 million people in order to be credible [38]. As this represents approximately 7.9% of the total population [38], an admittedly rough and unsubtle approximation of the expected value of credible deterrent for advanced civilizations might be ≈ 0.1 x *total loss*. Even lower values, such as approximately one million deaths and the destruction of sizable part of infrastructure, have been suggested [39]. We might conclude that credible deterrent may be achieved by being able to threaten 0.01…0.1 x *total loss*. As a real-life example, Chinese nuclear strategy is based on the assumption that deterrence is achieved by the capability to threaten only a few largest population and industrial centers in retaliation [40].

If just one interstellar probe slipping through the defenses could endanger the entire planet-bound part of the AC's civilization, it would seem that probabilities of counter-attack rising towards 0.2 would at the very least make would-be Attacking Civilizations cautious. As seen above, such probabilities arise even when the AC is extremely confident of its abilities.

## 5. Discussion and conclusions

In this paper, I have argued that any interstellar attack is a hazardous gamble for the attacker. It seems that if interstellar travel and warfare are at all possible, then any civilizations that have a reason to fear another civilization (i.e. are not so far advanced that another civilization simply cannot harm them) also have a reason to fear eventual retaliation if they attempt to strike first. In order to lower the probability of counter-attack to levels that are seen to be non-credible deterrents among humans, the Attacking Civilization would need to be *extremely* certain of being able to destroy the Victim Civilization (and its allies) and avoid preemptive attacks from $N$th Civilizations.

In particular, the time lag between detection and the arrival of the attacking force seems to pose extremely grave challenges for the attacker. Even under very optimistic assumptions for the density of civilizations in our galaxy, the victim civilization would seem to have a distinct chance of spreading out from its home planet and developing at least rudimentary survival and retaliatory capability. If the average distance between civilizations is short, on the other hand, the greatest threat to the attacking civilization may actually come from *N*th Civilizations: If the average distance between civilizations is short, on the other hand, the greatest threat to the attacking civilization may actually come from *N*th Civilizations: if the average distance between civilizations is so short that the Victim Civilizations are unable to develop effective defenses and retaliatory capabilities, the implication is that civilizations are common and therefore it is likely that a large number of civilizations will be in position to witness the aggression. Some of these may very well be far more advanced than the AC or be able to ally against the AC (or even be allies of the VC), and for reasons of self-protection, they may not wish to be neighbors with perpetrators of interstellar genocide, even if they cannot or will not attack the AC immediately. Although the NC will also risk retaliation if it launches a preemptive attack, it can be argued that risking retaliation when the adversary has demonstrated its willingness to strike first and revealed its capabilities, in contrast to an unprovoked preventive attack against unknown adversary, entails qualitatively different conceptions of risk. What probability of retaliation is required to deter an interstellar adversary is, of course, open to debate. It is, in fact, my clear intention to encourage the further development of the ideas expressed in this paper and the use of explicit methodologies to study quantitatively the risks of possible first contact. A follow-up to this paper will develop the simplified model towards a more complete model of interstellar conflict, including its analysis through computer simulation. Necessary areas of improvement include more detailed simulation of the dynamics of interstellar attack; in particular, taking into account spatio-temporal distributions of civilizations and their effect on the dynamics would be welcome.

The study also suffers from obvious limitations, the chief of which is that we cannot know much anything about the reasoning processes of possible ETIs. As plausible as it sounds to argue that any advanced civilizations must be somewhat rational in the sense we would understand it, the specifics of that rationality will certainly differ. Regarding the subject of this paper, it is almost equally plausible that the ETIs would consider the elimination of "inferior" species their sacred duty, to be undertaken despite the risks. Likewise, it is possible that extremely advanced "true" star faring civilizations will have their reasons to destroy less advanced civilizations even if they cannot pose a threat to them. It is also possible that a civilization will develop technologies that make them practically invulnerable to any retaliation from other civilizations in the vicinity.

Unfortunately, we cannot draw from better sources than our own history when arguing whether or not a civilization would initiate a preventive attack against other civilization. Based on that history, it would seem that any possibly paranoid ETIs that may be listening for humanity's signals are in a position that is roughly analogous to the position of the U.S. in the period between the end of the World War Two and the first Soviet nuclear bomb. Hostile ETIs could probably hit us severely and have a good chance of destroying us for good, but they must also assume that every year brings us closer to the capability to retaliate. If the human example is anything like the average, the possibility of receiving just one hit from a high-velocity interstellar probe on exchange should make the would-be attacker think twice before committing acts of aggression.

Of course, whether our capabilities would in the end be enough to deter attack is something that cannot be conclusively proved, except in the negative. However, if the current pace of technological development continues, the ETIs need to attack during this century – which would require them to have ready forces within approximately 100 light years, if detection is assumed to be based on leaked electromagnetic

radiation – or they will likely have to confront a civilization that already has a sizable presence in space and is building its first interstellar spacecraft. Reliably destroying such a civilization may prove to be challenging, perhaps challenging enough to cause them to reconsider.

However, the designers of interstellar spacecraft need to consider that their creations may be seen as threatening by other species. It is unfortunately all too easy to imagine a scenario where a human flyby probe to a supposedly uninhabited system accidentally damages a civilization that had chosen to remain quiet, perhaps due to paranoid fear of detection, and the said civilization sees no alternative but to strike back in order to stop further "attacks." In short, the mission planning of any interstellar spacecraft *must* ensure beyond reasonable doubt that the target system either does not host intelligent life or that the mission will not even appear to pose a danger to them. Of course, detecting intelligent life that does not want to be detected will be a difficult challenge.

The METI effort should view these arguments as reasons for cautious optimism. It would seem that if ETIs exist in our close neighborhood, any ETI whose technology is not far in advance of current human technology would also be likely to be deterred from attacking us. On the other hand, an ETI whose technological abilities make it invulnerable to whatever humanity can possibly devise would probably also be able to detect us, METI or no METI – but such ETIs would seem to have little reason to wish us harm. It can also be argued that hostile ETIs need to consider whether METI signals are representative of true technological developments or a lure designed to give the impression of weakness; this alone should serve as a deterrent of sorts against outright preventive attacks at the least.

## 6. Acknowledgements

The author would like to thank Markku Karhunen and two anonymous referees, whose valuable comments greatly improved this article, and Jenny and Antti Wihuri Foundation, whose generous grant provided time for the preparation of this study.